\documentclass[10pt,journal,compsoc]{IEEEtran}
\usepackage{dirtytalk}
\usepackage{amsmath}
\usepackage{amssymb}
\usepackage{amsfonts}
\usepackage{amstext}
\usepackage{amsthm}
\usepackage{comment}
\usepackage{caption}
\usepackage{subcaption}

\ifCLASSOPTIONcompsoc
  \usepackage[nocompress]{cite}
\else
  \usepackage{cite}
\fi
\ifCLASSINFOpdf
  \usepackage[pdftex]{graphicx}
  \graphicspath{{./Images/}}
  \DeclareGraphicsExtensions{.pdf,.jpeg,.png}
\else
  \usepackage[dvips]{graphicx}
  \graphicspath{{./Images/}}
  \DeclareGraphicsExtensions{.eps}
\fi
\usepackage{array}
\usepackage{url}
\begin{document}
\title{Implicit Feedback Deep Collaborative Filtering Product Recommendation System}

\author{Karthik Raja~Kalaiselvi Bhaskar,~\IEEEmembership{University of Toronto,}
        Deepa~Kundur,~\IEEEmembership{University of Toronto,}
        and~Yuri~Lawryshyn,~\IEEEmembership{University of Toronto}% <-this % stops a space
\IEEEcompsocitemizethanks{\IEEEcompsocthanksitem Karthik Raja Kalaiselvi Bhaskar is an M.A.Sc graduate with the Department of Electrical and Computer Engineering, University of Toronto.\protect\\
Email: kbhaskar@ece.utoronto.ca
\IEEEcompsocthanksitem Prof. Deepa Kundur is with the Department
of Electrical and Computer Engineering, University of Toronto.\protect\\
Email: dkundur@ece.utoronto.ca

\IEEEcompsocthanksitem Prof. Yuri Lawryshyn is with the Department
of Chemical Engineering, University of Toronto.\protect\\
Email: yuri.lawryshyn@utoronto.ca
}% <-this % stops an unwanted space
}

\IEEEtitleabstractindextext{%
\begin{abstract}
In this paper, several Collaborative Filtering (CF) approaches with latent variable methods were studied using user-item interactions to capture important hidden variations of the sparse customer purchasing behaviors. The latent factors are used to generalize the purchasing pattern of the customers and to provide product recommendations. CF with Neural Collaborative Filtering (NCF) was shown to produce the highest Normalized Discounted Cumulative Gain (NDCG) performance on the real-world proprietary dataset provided by a large parts supply company. Different hyperparameters were tested using Bayesian Optimization (BO) for applicability in the CF framework. External data sources like click-data and metrics like Clickthrough Rate (CTR) were reviewed for potential extensions to the work presented. The work shown in this paper provides techniques the Company can use to provide product recommendations to enhance revenues, attract new customers, and gain advantages over competitors.
\end{abstract}

}

\maketitle

\IEEEdisplaynontitleabstractindextext

% \fi
\IEEEpeerreviewmaketitle

\IEEEraisesectionheading{\section{Introduction}\label{sec:introduction}}

%\section{Introduction}

With today’s ever-increasing ease of access to the internet and information, we have reached a point of information overload. Users often find themselves in a dilemma where they spend more time sorting through the clutter of information rather than absorbing the information itself. Over the last decade, there has been a push towards creating accurate and efficient recommendation systems to help users sort through this clutter and improve the experience of obtaining information. However, with the recent increase in research in Machine Learning and Deep Learning, researchers have been experimenting with using deep learning techniques to provide recommendations rather than relying on traditional recommendation techniques like probabilistic matrix factorization \cite{mnih2008probabilistic} and Naive Bayes classifier \cite{aiguzhinov2010similarity}.

The recent emergence of information overload has caused decreased productivity, increased frustration, and an overall more negative user experience \cite{zou2014information}. For example, users now often spend more time trying to find a movie on Netflix to watch rather than actually watching movies \cite{gomez2016netflix}.

As with any industry, technology companies ultimately focus on their revenues and profits as key metrics for optimization. Recommendation systems play a significant role in increasing revenues, retaining customers and users, and gaining competitive advantages over competitors \cite{recsys}. By using recommendation systems, companies can encourage users to stay engaged with their application/website/service, which allows the companies to show more advertisements, attract new clients, and retain existing clients \cite{vaidya2017recommender}. From an e-commerce perspective, recommendation systems can entice customers to buy more. Carefully selected recommendations can make customers more satisfied, leading them to buy more items than they previously anticipated.

Companies store large amounts of data in databases containing transactional, accounting, inventory, and customer contact information. These data can be analyzed to understand customer purchasing behavior, which allows companies to target customers through product recommendations based on past purchases or activities leading to increased revenue. Companies such as Amazon and Netflix are already using algorithms to give recommendations, and other companies are looking to implement similar recommendation system algorithms. The development of e-commerce solutions pioneered by Amazon not only demonstrated the importance of incorporating web data to meet customers’ shopping needs but also leveraging customer data to enhance sales using machine learning algorithms. In another example, Uber Eats food recommendations are based on many factors that take into account real-time information such as user queries, location, time of day/week and historical information about the user’s purchases, restaurants and delivery partners used \cite{uber}.

A large parts supply company (the Company) has gathered a large amount of sales and transactional information and has an extensive product database. However, the Company performed minimal analysis of their databases from the perspective of product recommendations to customers. Furthermore, the Company’s databases are large, containing more than 500,000 stock keeping units (SKUs), 20 million rows of transactional data as well as information related to their 200,000 customers. The two primary databases are Invoiced Orders (Invoicedorders.csv) and Current Items (ITEMCURRENT.csv).  The \say{Invoiced Orders} database contains the customers’ transaction  data.   The  \say{Current  Items}  database  contains  detailed  information  about  the products offered by the Company. The Company wants to use the available data to provide a list of 12 products the customers may be interested in purchasing and to improve the user experience of the Company's website.

Collaborative Filtering (CF) was used to solve this problem, it relies on the intuitive idea that similar users tend to like different items similarly. CF methods use an a-priori available set of user-item ratings to learn the interdependencies among users and items, and then predict a user’s rating of an item either via the neighboring items’ ratings (neighbor-based \cite{das2007google}, \cite{sarwar2001item} ) or by inferring latent factors that find similar users and items in a low-dimensional embedding (latent factor-based  \cite{hu2008collaborative}, \cite{he2017neural}, \cite{rendle2009bpr}, \cite{moussawi2018towards} ). The feedback can be explicit, i.e., the user either likes an item or dislikes an item. For example, the user either clicks “thumbs up” or “thumbs down” on a youtube video or rates a movie on a scale of one to five stars. Alternatively, the feedback can be implicit, i.e., the feedback derived from the browsing behavior of the user e.g., the user clicked on/purchased a product, checked into a venue, or viewed an article.

This paper aims to apply and build a state of the art recommender system algorithms on a real-world dataset. We used machine learning algorithms on the Company’s transactional database, which will allow the Company to predict their customers’ needs better and provide product recommendations tailored to the customers’ purchasing behaviours. The contributions associated with this work are listed below:
\begin{itemize}%[\IEEEsetlabelwidth{Z}]
%\item  Pre-processed the Invoiced Orders and Current Items databases from the Company and made the data suitable for the recommendation system problem. 
\item The evaluation and testing of the hypothesis that Collaborative Filtering with latent variable models can capture the historical, transactional and sparse customer purchasing behaviors to provide personalized product recommendations. This paper work continues related efforts in \cite{hu2008collaborative}, \cite{he2017neural}, \cite{rendle2009bpr}, \cite{moussawi2018towards} to understand the hidden factors in customers’ purchasing behaviors.
\item The construction and evaluation of ranking metrics for four different models based on Matrix Factorization using Alternative Least Squares, Bayesian Personalized Ranking, Neural Collaborative Filtering and Autoencdoer Collaborative Filtering.
\item Applied Bayesian Optimization (BO) for hyperparameter tuning. We used BO to evaluate the next set of parameters using previous hyperparameters, reducing the cost of searching the hyperparameter space.
\end{itemize}

\section{Related Work}
%\IEEEraisesectionheading{\section{Related Work}\label{sec:relatedwork}}

A recommendation system, often also called a \say{recommender system}, is used to estimate user preferences of items or objects they have not seen yet. Recommendation systems commonly use inputs such as user preferences, item/object features, histories of users-items’ past interactions, temporal data, and spatial data. Generally, there are three types of recommendation systems: Collaborative Filtering systems  \cite{ekstrand2011collaborative}, \cite{zhao2010user}, \cite{sarwar2001item} , Content-Based Recommendation systems \cite{zenebe2009representation}, \cite{lops2011content}, and Hybrid systems\cite{burke2002hybrid},  \cite{burke2007hybrid}. Latent-factor or matrix factorization (MF) methods are popular in recommendation systems  \cite{koren2009matrix}  both for implicit \cite{rendle2009bpr},\cite{hu2008collaborative} and explicit feedback \cite{mnih2008probabilistic}. The latent factor model tries to predict the users' rating on an item by optimizing an objective function and by reconstructing the rating matrix into  low-rank dimensional latent factors. Bayesian Personalized Ranking (BPR) \cite{rendle2009bpr} has emerged as one of the best Top-K recommendation models for implicit data. Popular deep learning-based recommendation techniques that utilize multi-layer perceptrons are Neural Collaborative Filtering (NCF) \cite{he2017neural} and deep factorization machine \cite{guo2017deepfm}. In MLP techniques, recommendations are considered as two-way interactions between users’ preferences and the features of an item/object. NCF uses the binary cross-entropy loss function for implicit feedback and the weighted square loss for explicit feedback. Further extension of the work using pairwise ranking loss are proposed by the authors in \cite{niu2018neural}, \cite{song2018neural} to improve the results. The NCF model is extended to cross-domain recommendations like \cite{lian2017cccfnet}, \cite{wang2016collaborative}. DeepFM \cite{guo2017deepfm} is an end-to-end model, able to integrate factorization machine and MLP to find lower and higher-order feature representation. DeepFM is similar to the wide and deep model, a two network deep learning architecture gives recommendations  \cite{cheng2016wide}; however, it does not require sophisticated feature engineering. The authors in \cite{lian2018xdeepfm} proposed eXtreme deep factorization machine, which models implicit and explicit features together to improve the performance over DeepFM. Covington et al. proposed an MLP based YouTube recommendation model \cite{covington2016deep}, and the predictor generates a top-n list of videos based on the nearest neighbor scores from the several hundred videos. The authors in \cite{alashkar2017examples} explored and applied MLP in makeup recommendations. This work uses two identical MLPs to model labeled examples and expert rules, respectively, which provides highly precise recommendations.

Autoencoders are also widely used  in building recommendation systems. There are two general techniques for using autoencoders as recommendation systems. First, the autoencoder can be used to learn a lower-dimensional feature representation at the bottleneck layer, which means the employement of autoencoder as a dimensionality reduction tool, which is used sequentially with other deep learning techniques for recommendations  \cite{zhang2019deep}. On the other hand, one can use autoencoders to fill in the blank values of the rating matrix directly in the reconstruction or decoder layer \cite{zhang2019deep}. Nearly all of the autoencoder alternatives, such as denoising autoencoder, variational autoencoder, contactive autoencoder, and marginalized autoencoder, can be employed to the recommendation task \cite{zhang2019deep}. The AutoRec \cite{sedhain2015autorec} system is a specific implementation of the technique to use the autoencoder as a generative tool for recommendations. The Collaborative Filtering Neural Network \cite{strub2015collaborative}, \cite{strub2016hybrid} is an extension of the AutoRec System. It uses denoising techniques to make the recommendation system more robust. It additionally uses side information (user profile information / item descriptions) to reduce data sparsity issues and the cold start problem, which in return, increases the training speed and robustness, and improves the prediction accuracy. Unlike  other autoencoders, which are designed to output rating predictions, Collaborative Denoising Autoencoders (CDAE) models \cite{wu2016collaborative} are used to output ranking predictions and are prone to overfitting, because CDAE provides multiple predictions at once; for example, “What are the top 10 best movies for User A”. The authors in \cite{liang2018variational} proposed a modification of the variational autoencoder called Multi-VAE and Multi-DAE for recommendation tasks using implicit data. The proposed alternative showed better performance than CDAE.

Matrix Factorization with Alternating Least Square (ALS) \cite{hu2008collaborative}, Bayesian Personalized Ranking (BPR) \cite{rendle2009bpr}, Neural Collaborative Filtering (NCF) \cite{he2017neural} and Autoencoder for Collaborative Filtering (ACF) \cite{moussawi2018towards} are the four different latent variable methods that were selected to analyze large customer purchasing behavior patterns because of the implicit feedback between customers and products. These modelling approaches were selected to examine different aspects of providing product recommendations.

\section{Methodology}

The specific details of the dataset cannot be directly summarized in this paper since the data are proprietary. The two primary databases are Invoiced Orders (Invoiced\_orders.csv) and Current Items (ITEM\_CURRENT.csv).

 The \say{Invoiced Orders} database contains 20 million historical transactional data from August 2016 to June 2019. It has more than 100 fields such as geographic location, selling location, regional order number, rebate value,
stock quantity, etc. The \say{Current Items} database contains 500,000 unique products and has detailed information about the items offered by the Company, including dimensions, quantity, English, and French text descriptions, etc. 

The focus of the analysis was to recommend products/items based on past transaction history; therefore, most of the information is filtered out like location, manufacturer, order\_date, etc. The dataset is filtered using a structured query language (SQL) query where the final dataset has the following information: customer ID, product ID, English and French descriptions of the products, and the number of times a customer purchased a product (ratings). 

\subsection{Matrix Factorization with Alternating Least Squares (ALS)}

Latent-factor or Matrix Factorization (MF) \cite{hu2008collaborative} methods are popular in recommendation systems, MF algorithms work by decomposing the user-item interaction matrix into the product of two lower dimensionality rectangular matrices. The latent factors (lower dimensions) otherwise called features, can be found using Singular Value Decomposition (SVD).  Let $M$ and $N$ denote the number of users and items, repectively and $k$ denotes the dimensions of latent space. Let $p \in \mathbb{R}^{k \times M} $ be the latent factor matrix for the users, where the $u^{th}$ column $p_u \in \mathbb{R}^k$ is the latent factor for user $u$. Similarly, let $q \in \mathbb{R}^{k \times N} $ be the latent factor for the items, where the $i^{th}$ column $q_i \in \mathbb{R}^k$ is the latent factor for item $i$. The ratings ($\hat{r}_{u, i}$) for user $u$ and item $i$ are given by $\hat{r}_{u, i} = q_i^T p_u$. For implicit rating, the MF can be formulated \cite{hu2008collaborative} as
\begin{equation}
 min \sum c_{(u, i)} ( p_{(u, i)} - q_i^T p_u)^2 + \lambda ( (\|q_i\| )^2 + (\|p_u\| )^2 )   
\end{equation}
where $c_{(u,i)} = 1 + \alpha r_{(u, i)}$ and $p_{(u, i)} = 1$ if $r_{(u, i)}>0$ and $p_{(u, i)} = 0$ if $r_{(u, i)} = 0$. $r_{(u, i)}$ is a numerical representation of users' preferences (e.g., number of purchases, number of clicks etc.) and $\lambda$ is a regularization parameter. Owing to the term of $q_i^T p_u$, the loss function is non-convex. The gradient descent method can be applied, but this will incur expensive computations. An Alternating Least Square (ALS) algorithm was therefore developed to overcome this issue \cite{hastie2015matrix}. The basic idea of ALS is to learn one of $q$ or $p$ at the time of optimization while keeping the other as constant. ALS makes the objective at each iteration convex and solvable. The alternating between $q$ and $p$ stops when the convergence is optimal. It is worth noting that the ALS iterative computation can be parallelized and/or distributed, which makes the algorithm desirable \cite{yu2014parallel} for use cases where the dataset is large and thus the user-item rating matrix is highly sparse (as typical in recommendation system scenarios).

\subsection{Bayesian Personalized Ranking (BPR)}
 BPR \cite{rendle2009bpr} has emerged as one of the best Top-K recommendation models for implicit data. BPR is also a strong baseline, which makes it difficult to beat. BPR falls under the category of one-class collaborative filtering (for example 0 or 1) and pairwise comparison. It considers the recommendation task as a ranking problem and assumes that the user prefers items that they have already observed/interacted with, rather than unobserved/not interacted with items. To learn the relative ranking of items for each user, BPR needs to model negative feedback. BPR uses the pairwise interpretation of positive-only feedback by creating triplets (user, observed item, unobserved item). The positive-only feedback is then transformed into positive and negative feedback in pairs of item $(i,j)$. $D_s$ represents triplets $(u,i,j)$ such that a user $u$ prefers item $i$ over item $j$. The triplets ($D_s$) are sampled \cite{rendle2009bpr} from
\begin{equation}
    D_s := (u,i,j)|i \in I_u^+ \land j \in I\backslash I_u^+
\end{equation}
where we let $U$ be the set of all users and $I$ the set of all items, with  implicit feedback $S$ defined as $ S \subseteq U \times I$; $I_u^+$ is defined as the items $I_u^+ := { i \in I : (u , i) \in S } $ where user $u$ gives positive feedback; $i$ is a positive item taken from $I_u^+$ and $j$ is a negative item randomly sampled from unobserved/not interacted with items.   The author derives an optimization criterion called BPR-OPT \cite{rendle2009bpr}, which is an optimization framework. To provide a personalized ranking of items or recommendations, we need to train a separate model using BPR-OPT. Compared to standard MF or kNN methods, BPR ensures not only the rating predictions but also optimizes the item rankings.

\subsection{Neural Collaborative Filtering (NCF)}
\begin{figure}
     \centering

         \includegraphics[width=0.49\textwidth]{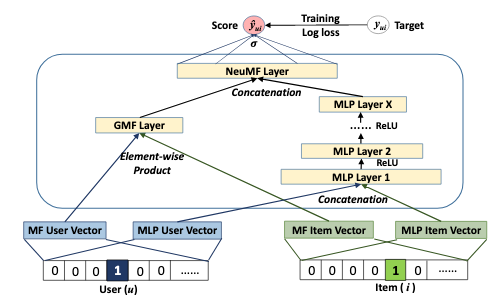}

        \caption{Neural Collaborative Filtering model architecture}
        \label{fig:ncf}
\end{figure}
NCF \cite{he2017neural} is a new neural matrix factorization model that combines both Generalized Matrix Factorization (GMF) and Multi-Layer Perceptron (MLP) to combine the strengths of the linearity and non-linearity given by MF and MLP, respectively, for modeling user-item latent features. The input layer consists of latent vectors of items and users. Figure \ref{fig:ncf} shows the architecture of NCF. The User (u) and Item (i) are used to create low-dimensional embeddings for the user and item. Generalized Matrix Factorization (GMF) combines the two embeddings using the dot product. Multi-layer perceptron (MLP) also produces embeddings for the user and items. However, instead of taking a dot product to obtain the rating, the embeddings are concatenated to create a feature vector that can be passed on to the further layers. The outputs from the final layers of the MLP  and GMF are concatenated, called a NeuMF layer to obtain the prediction score. The final layer output of GMF ($\hat{r}_{u,i}$) can be formulated as follows:
\begin{equation}
    \hat{r}_{u,i} = a_{out} ( h^T (q_i \odot p_u ) )
\end{equation}
where we let the user latent vector $p_u$ be $P^T v_u^U$ and item latent vector $q_i$ be $Q^T  v_i^I$, where $ \odot $ is an element-wise product of vector terms. Additionally, $a_{out}$ and $h$ represent the activation function and weights of the output layer, respectively. The MLP model under the NCF framework is defined as follows. In the input layer, we  concatenate the user latent vector $p_u$ and item latent vector $q_i$ as follows:
\begin{equation}
    z_1 = \phi_1 (p_u, q_i) = \begin{bmatrix} p_u \\ q_i\end{bmatrix}
\end{equation}
where $z_1$ represents the concatenation of $p_u$ and $q_i$ at the input layer. The hidden layers are formulated as:
\begin{equation}
    \phi_l(z_l) = a_{out} (W_l^T z_l + b_l) , (l = 2, 3,  ..., L-1)
\end{equation}
where $W_l$, $b_l$, and $a_{out}$ denote the weight matrix, bias vector, and activation function for the $l$-th layer's perceptron, respectively and the output layer is formulated as:
\begin{equation}
    \hat{r}_{u,i} = \sigma ( h^T \phi(z_{L-1}))
\end{equation}
where $\hat{r}_{u,i}$ is the predicted rating, $h$ denotes the edge weights and the activation function (sigmoid) of the output layer is defined as $\sigma (x) = \frac{1}{1 + \mathrm{e}^{-x} }$ to restrict the predicted score to be in (0,1). To have more flexibility in the fused model, we use GMF and MLP to learn the  embeddings separately and then combine these two models by concatenating their last hidden layer \cite{he2017neural}. We get $\phi^{GMF}$ from GMF and obtain $\phi^{MLP}$ from MLP:
\begin{equation}
    \phi^{GMF}_{u, i} = p^{GMF}_u \odot q^{GMF}_i
\end{equation}
\begin{equation}
    \phi^{MLP}_{u,i} = a_{out} \big( W_L^T \big( a_{out} \big( .. a_{out} \big( W_2^T \begin{bmatrix} p_u^{MLP} \\ q_i^{MLP}\end{bmatrix} + b_2 \big) ..\big) \big) + b_L \big)
\end{equation}
Lastly, we fuse the output from GMF and MLP:
\begin{equation}
    \hat{r}_{u,i} = \sigma \bigg( h^T \begin{bmatrix} \phi^{GMF} \\ \phi^{MLP}\end{bmatrix} \bigg)
\end{equation}
where $\hat{r}_{u,i}$, $\sigma$, $h$, $\phi^{GMF}$ and $\phi^{MLP}$ denote the predicted ratings, sigmoid activation function, edge weights of the output layer, last hidden layer of GMF and last hidden layer of MLP, respectively. By taking a negative log likelihood, we obtain the objective function to minimize for the NCF method:
\begin{equation}
    L = - \sum_{ (u,i) \in \mathbb{O} \cup \mathbb{O}^{-}} r_{u,i} log \hat{r}_{u,i} + ( 1 - r_{u,i} ) log ( 1 - \hat{r}_{u,i} )
\end{equation}
where $\mathbb{O}$ denotes the set of observed interactions, $\mathbb{O}^{-}$ denotes the set of negative instances (unobserved interactions), $r_{u,i}$ denotes the actual ratings and $\hat{r}_{u,i}$ denotes the predicted ratings.

\subsection{Autoencoder for Collaborative Filtering (ACF)}
Recently, there has been a significant focus in research on using autoencoders for recommendation systems  \cite{sedhain2015autorec}, \cite{strub2015collaborative}. Many different techniques have been proposed, such as denoising architecture  \cite{wu2016collaborative}, dropout to increase efficiency, etc. Here, we use a particular type of autoencoder (ACF) \cite{moussawi2018towards} whose model is described in this section. Consider the user-item interaction matrix is represented as $ X \in \{0,1 \}^{|U| \times |I|} $ where $U$ and $I$ are the set of users and items, respectively. If there is an interaction between user $u$ and item $i$ then $X_{u,i} = 1$; otherwise,  $X_{u,i} = 0$. Given user $u$ and item $i$, $I_u$ represents a set of items, $u$ has interacted with, and $U_i$ represents a set of users who have interacted with $i$. The autoencoder learns a model $p(x_u | z_u , \theta) = h(g_\theta(z_u))$, where $x_u$ is the user $u$ vector of interactions, $z_u$ represents the user latent factor, $g_\theta$ is an autoencoder parameterized by $\theta$ and $h$ is an activation function that maps the output of $g_\theta$ to probabilities based on the logistic likelihood distribution used to model $p(x_u | z_u , \theta)$. $z_u$ can be computed as a function of $f_\lambda(x_u)$, where $f_\lambda$ is an autoencoder parameterized by $\lambda$. The negative log-likelihood loss function of our model to be minimized is then:
\begin{eqnarray}
    - \sum_i \log p(x_u | z_u, \theta)_i = - x_u \cdot \log (g_\theta ( z_u)) \nonumber\\
    - ( 1 - x_u) \cdot \log (1 - g_\theta ( z_u))
\end{eqnarray}
For regularization, dropouts are applied at the input layers, and also applied L2 weight decay on $\theta$ and $\lambda$.

\section{Evaluation and Model Specifications}

\subsection{Evaluation Protocols}

To evaluate the efficiency of a product recommendation, we followed the leave-one-out evaluation \cite{he2017neural}, which has been widely used in the literature \cite{bayer2017generic}, \cite{he2016fast}, \cite{rendle2012bpr}. In leave-one-out evaluation, for each user, we keep their latest interaction with the item as a test set, and we use the remaining as a training set for the model. During evaluation, it is too tedious and time-consuming to rank all the items for each user. To overcome this problem, we adopted a common strategy used by the practitioners where we sample 100 random items that are not interacted with by the user as a test set and then rank those 100 items with respect to that user. Leave-one-out evaluation with negative sampling is implemented in this  work. The efficiency of the ranked list is measured by a widely used metric in the learning-to-rank problem called Normalized Discounted Cumulative Gain (NDCG) \cite{yilmaz2008simple}. NDCG gives more weight to relevant predictions at the starting of the ranked list of items and discounts relevant predictions that occur farther from the beginning of the ranked list. One-product Hit Ratio is a metric we devised, which is relevant for the business. One-product Hit Ratio intuitively measures whether the test item is present on the ranked list. For example, if the test item present on the ranked list, then the score is 1; otherwise 0.

\subsection{Model Specifications}

The model specifications include the model training procedures and selection of hyperparameters for the Collaborative Filtering models. These models are prepared using Tensorflow \cite{abadi2016tensorflow} and PyTorch \cite{paszke2017automatic} frameworks with the Python programming language. These models are trained on a server with 32 cores of CPU and 256 GB of RAM. The dataset is split into a train, validation and a test set using a customized function, where the split is stratified so that the same set of customers and products will appear in all train, validation and test sets. Hyperparameter tuning is an essential part in machine learning; grid search, random search and bayesian optimization are the three standard methods for hyperparameter tuning. The Bayesian Optimization approach has been proven to outperform other state-of-the-art hyperparameter tuning approaches \cite{snoek2012practical}. Bayesian approach uses past choices made to make a smart choice of hyperparameters for the next set of values to evaluate, through which it reduces the cost of searching for parameters. In this paper, we use Bayesian Optimization for hyperparameter tuning.

For the ALS methodology, we closely followed the implementation of \cite{hu2008collaborative} as mentioned in Section 3.1. We find that training the model in this fashion takes more time, so we experimented with different optimizers and used the Adam optimizer \cite{kingma2014adam} to reduce the training time. To make the training even faster, we converted the python code to Cython \cite{behnel2011cython}, where Cython converts the python code to C code to boost performance. We used Tensorflow's embedding layer to create user-item latent vectors. We used Bayesian Optimization to find the optimal hyperparameters. Table \ref{tab:MF_ALS} displays the hyperparameters of the ALS model. The hyperparameters include user-item latent dimensions, regularization, iterations and scaling factors. We find that the user-item latent dimensions have the most significant impact on performance. To create latent dimension vectors, we used embedding layers whose values were uniform and randomly initialized, and then the values were learned during the model training process. L2 regularization was used. The scaling factor was set to 15. During hyperparameter tuning, the number of iterations was found to be optimal at 50.

\begin{comment}
\begin{figure*}[!t]
\centering
\subfloat[Case I]{\includegraphics[width=2.5in]{Images/chap4/ALS_TL_k_200_reg_le4.png}
\label{fig_first_case}}
\hfil
\subfloat[Case II]{\includegraphics[width=2.5in]{Images/chap4/BPR_TL_k_200_reg_le4.png}
\label{fig_second_case}}
\caption{Simulation results for the network.}
\label{fig_sim}
\end{figure*}

\begin{figure}[!t]
\centering
\includegraphics[width=2.5in]{Images/chap4/ALS_TL_k_200_reg_le4.png}
\caption{MF-ALS Training Loss Graph}
\label{fig_sim}
\end{figure}
\end{comment}
\begin{table}[ht]
\caption{Hyperparameters for MF with ALS models}
\begin{center}
\begin{tabular}{ |c|c|c|c| } 
\hline
\textbf{Hyperparameter} & \textbf{Value}  \\
\hline
user-item Latent Dimension (k) & 200 \\
Regularization & 0.0001 \\
Learning Rate & 0.01 \\
Iterations (Epochs) & 30 \\
Batch Size & 512 \\
Scaling Factor ($\alpha$) & 15 \\
\hline
\end{tabular}
\end{center}
\label{tab:MF_ALS}
\end{table}

In MF with the ALS method described above, we focused on pointwise loss minimization, which captured the positive or negative user-item interactions separately. However, there may be some hidden information available in the negative user-item interactions for positive user-item interactions. The idea was formulated into pairwise loss minimization by \cite{rendle2009bpr}. We train the triplets using the pairwise loss function, as discussed in Section 3.5. The BPR latent factors are constructed using TensorFlow’s embedding layer, and we find that BPR has the same training characteristics as MF with the ALS method. However, BPR takes more iterations or epochs to converge to local minima. This problem arises because of the pairwise loss function, which needs to compute more gradients during the training procedure. The hyperparameter for this methodology consists of user-item latent dimensions (factors), the number of epochs, the learning rate for the optimizer, and regularization to prevent overfitting. By using Bayesian Optimization, we find the best hyperparameters for the BPR approach. Table \ref{tab:BPR} displays the hyperparameters of the BPR model.

\begin{table}[ht]
\caption{Hyperparameters for BPR model}
\begin{center}
\begin{tabular}{ |c|c|c|c| } 
\hline
\textbf{Hyperparameter} & \textbf{Value}  \\
\hline
user-item Latent Dimension (k) & 200 \\
Learning Rate & 0.01 \\
Regularization & 0.0001 \\
Iterations (Epochs) & 200 \\
Batch Size & 512\\
\hline
\end{tabular}
\end{center}
\label{tab:BPR}
\end{table}

In ALS and BPR methods, we tried to find the linear relationships that exist between the user and the item. However, in NCF, we try to find both linear and non-linear relationships in user-item interactions by utilizing the power of neural networks. As discussed in Section 3.3, we create GMF, MLP, and Fusion of GMF \& MLP (NeuMF) models. The goal of the NCF model is to train and minimize the binary cross-entropy loss function as defined in equation (10). The hyperparameters for the NCF methodology consist of n\_factors, layer\_size, n\_epochs, learning\_rate, and batch\_size, where n\_factors represents the dimensions of the latent space; layer\_size represents the sizes of the input and hidden layers of the MLP; n\_epochs is the number of iterations to run the training. In general, we find that increasing n\_factors increases the quality of predictions. The user/item labels are mapped to real-valued latent vectors with Tensorflow's embedding layers. Table \ref{tab:NCF} describes the optimal hyperparameters for the NCF models using Bayesian Optimization. We found that, in training a NeuMF model, using pre-trained model weights of GMF and MLP is far better in reducing cross-entropy loss than using gaussian normal sampled initialized weights.

\begin{table}[ht]
\caption{Hyperparameters for NCF model}
\begin{center}
\begin{tabular}{ |c|c|c|c| } 
\hline
\textbf{Hyperparameter} & \textbf{Value}  \\
\hline
n\_factors & 16 \\
layer\_size & [64,32,16] \\
n\_epochs & 50 \\
learning\_rate & 0.001 \\
batch\_size & 256 \\
\hline
\end{tabular}
\end{center}
\label{tab:NCF}
\end{table}

The ACF approach is entirely different from previous approaches. Here, we use the Autoencoder both as a tool for dimensionality reduction as well as a learning algorithm to discover the hidden user-item latent features in the dataset. We use the PyTorch framework to build the ACF model. The objective of the ACF model is to minimize the loss function which was defined in equation (11). Typically, an Autoencoder has two parts: an encoder and a decoder. PyTorch’s nn.embedding layers are used to build the encoder, whereas the same nn.embedding layers are used to build the decoder, but we reverse the encoder’s architecture. User/item labels are mapped to the latent space using an encoder. As the name suggests, the decoder is used to decode the encoder to get the original user/item labels from the latent space. Instead of randomly assigning weights to the embedding layer, Xavier’s initialization is used. An Adam optimizer, as well as the ReLU activation function, are used throughout the process.  The hyperparameter for the ACF model consists of hidden\_layer, noise\_prob (dropout probability at the input layer), dropout\_prob (dropout probability at the bottleneck layer), lr (learning rate), weight\_decay, batch\_size, and  num\_epochs. Like ALS, BPR and NCF methodologies, we use Bayesian Optimization to find the best hyperparameters. Table \ref{tab:ACF} shows the hyperparameters for the ACF model.

\begin{table}[ht]
\caption{Hyperparameters for ACF model}
\begin{center}
\begin{tabular}{ |c|c|c|c| } 
\hline
\textbf{Hyperparameter} & \textbf{Value}  \\
\hline
hidden\_layer & 7 \\
noise\_prob & 0.3 \\
dropout\_prob & 0.2 \\
lr (learning rate) & 0.001 \\
weight\_decay & 2e-5 \\
batch\_size & 256 \\
num\_epochs & 30 \\
\hline
\end{tabular}
\end{center}
\label{tab:ACF}
\end{table}

\section{Results and Discussion}
\subsection{Results}

The average performance of the CF model predictions for the test set using leave-one-out evaluation are displayed in Table \ref{tab:Results}. The NDCG metric was used to calculate the performance of the model, according to the business objective, we choose twelve products to recommend; hence, we used NDCG@12. NDCG@12 gives us the twelve most relevant ranked products based on customer purchasing behavior. The results in the Table \ref{tab:Results} demonstrate that the NCF model achieved the best average performance for NDCG@12 over the leave-one-out evaluation test interval. One-Product Hit Ratio is 1 for ALS, BPR, NCF, and ACF; which means the algorithms at least predict one product that the customer intend to buy.

\begin{table}[ht]
\caption{CF average performance metrics}
\begin{center}
\begin{tabular}{ |c|c|c|c| } 
\hline
\textbf{Average Predictions} & \textbf{NDCG@12} & \textbf{One-Product Hit Ratio} \\
\hline
ALS & 0.577 $\pm$ 0.055 & 1 \\
\hline
BPR & 0.636 $\pm$ 0.048 & 1\\
\hline
NCF & 0.724 $\pm$ 0.049 & 1\\
\hline
ACF & 0.604 $\pm$ 0.069 & 1\\
\hline
\end{tabular}
\end{center}
\label{tab:Results}
\end{table}

We consider ALS as the baseline since it tries to find the most direct relationship between the user and item using the matrix factorization technique. From the results in the Table \ref{tab:Results}, the performance of ALS is good over the test period. BPR performed better than ALS, which was expected, because it also implemented matrix factorization with the same user/item latent factors to discover the hidden features in the dataset. BPR varies from ALS in that we implemented pairwise training between positive and negative interactions, which was described in Section 3.2. NCF performed better than all other models, which was expected since it combines the linearity and non-linearity of matrix factorization and neural networks, respectively, which allows NCF to find more relevant hidden features over the dataset. The ACF performed better than ALS but not better than BPR and NCF, which is not always the case. The efficiency of an autoencoder depends on the data; it learns to capture as much information as possible rather than relevant information \cite{ghasemi2018neural}. This means that if we have less user-item interactions, the autoencoder may lose this information.

\subsection{Discussion}

For the specific dataset investigated in this paper, CF with NCF produced the best test performance. MF with ALS is considered as a baseline in this paper since it focused on direct linear relationships between a user and an item. From the results shown in Table 5.1, we can see that all the models performed relatively well compared with the baseline model. NCF has a significant 14.7\% increase in NDCG over the baseline and BPR has a  5.9\% increase in NDCG. Compared with BPR and NCF, ACF performance is relatively less, with a 2.7\% increase in NDCG over the baseline. The BPR matrix factorization model achieved better NDCG performance than the MF with the ALS model (the baseline), which indicates that the pairwise training in BPR was effective in exploring the parameter space as opposed to the pointwise training used in ALS, leading to the improved performance of NDCG, though both BPR and ALS used matrix factorization with the same user/item latent factor vector dimensions. The ACF did not perform as well as the BPR approach in terms of NDCG metrics. The huge sparsity in the dataset investigated likely limited the usefulness of the autoencoder-based deep learning approaches. As discussed before, the autoencoder captures more  irrelevant information as opposed to relevant information; since the dataset is sparse in nature, it is likely to capture non-relevant information on the user-item interactions. Although the autoencoder can be used as a dimensionality reduction tool, the sparsity issue does not allow us to use the autoencoder to its full potential. The NCF model combines matrix factorization and neural networks at the final layer, which allows the model to create a more effective and useful hidden representation of the sparse dataset that allows the model to perform better than other models.

\begin{figure}
     \centering
     \begin{subfigure}[b]{0.24\textwidth}
         \centering
         \includegraphics[width=\textwidth]{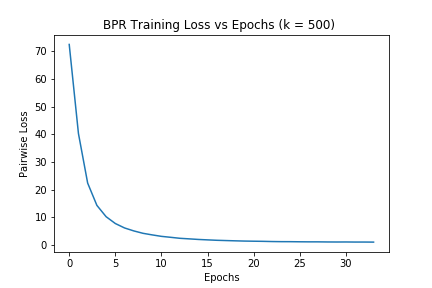}
         \caption{k = 500}
         \label{fig:BPR500}
     \end{subfigure}
     \hfill
     \begin{subfigure}[b]{0.24\textwidth}
         \centering
         \includegraphics[width=\textwidth]{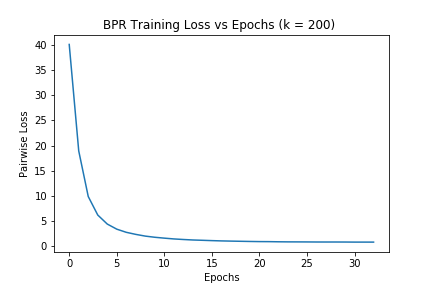}
         \caption{k = 200}
         \label{fig:BPR200}
     \end{subfigure}
        \caption{BPR Training Loss Graph}
        \label{fig:BPRK}
\end{figure}

\begin{figure}
     \centering
     \begin{subfigure}[b]{0.24\textwidth}
         \centering
         \includegraphics[width=\textwidth]{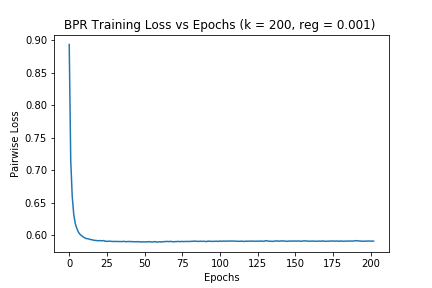}
         \caption{k = 200, reg = 0.001}
         \label{fig:BPRKR1}
     \end{subfigure}
     \hfill
     \begin{subfigure}[b]{0.24\textwidth}
         \centering
         \includegraphics[width=\textwidth]{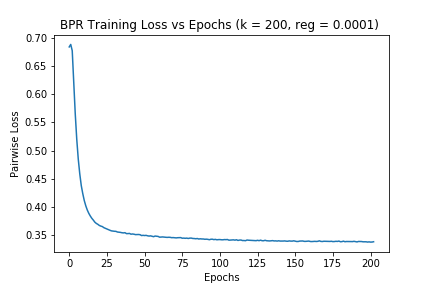}
         \caption{k = 200, reg = 0.0001}
         \label{fig:BPRKR2}
     \end{subfigure}
        \caption{BPR Training Loss Graph}
        \label{fig:BPRKR}
\end{figure}

Since the factors determined by the models are hidden, it is difficult to understand which factors of the user/item interaction are driving the results. We know that the hyperparameters of the model greatly affect performance. For the BPR model, the major hyperparameters that drive the results are user-item latent dimensional factors (k) and regularization. From Figure \ref{fig:BPR500}, we can see that for higher k values, the loss is relatively higher in the first few epochs compared with lower k values (Figure \ref{fig:BPR200}) by keeping other hyperparameters the same. Also, the model takes a significantly higher time to train when the k value is large. Another important hyperparameter is regularization; this is a technique used for tuning the loss function by adding an additional penalty term. From Figure \ref{fig:BPRKR1}, we can see that the pairwise loss is starting around 0.90 for a regularization value of 0.001. From Figure \ref{fig:BPRKR2}, it can be seen that the pairwise loss starts at 0.68 for a regularization value of 0.0001. Therefore, it is apparent that regularization is a driving factor in reducing the loss.

\begin{figure}
     \centering
     \begin{subfigure}[b]{0.24\textwidth}
         \centering
         \includegraphics[width=\textwidth]{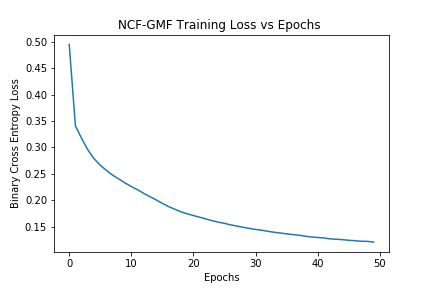}
         \caption{GMF}
         \label{fig:GMF}
     \end{subfigure}
     \hfill
     \begin{subfigure}[b]{0.24\textwidth}
         \centering
         \includegraphics[width=\textwidth]{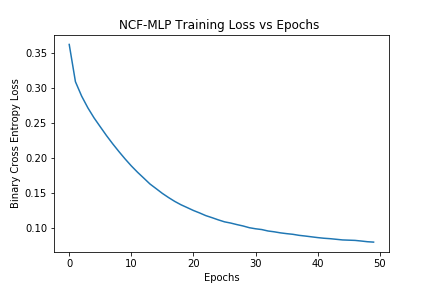}
         \caption{MLP}
         \label{fig:MLP}
     \end{subfigure}
        \caption{Training Loss Graph}
        \label{fig:NCF1}
\end{figure}

\begin{figure}
     \centering
     \begin{subfigure}[b]{0.24\textwidth}
         \centering
         \includegraphics[width=\textwidth]{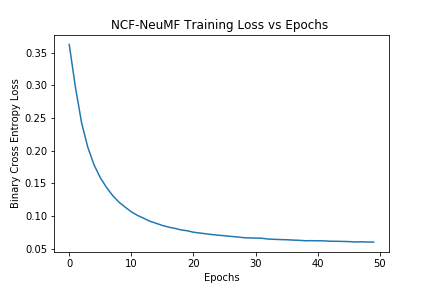}
         \caption{Without Pre-training}
         \label{fig:WPT}
     \end{subfigure}
     \hfill
     \begin{subfigure}[b]{0.24\textwidth}
         \centering
         \includegraphics[width=\textwidth]{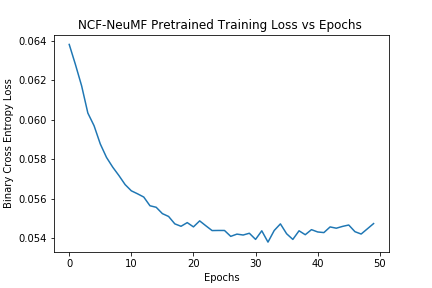}
         \caption{With Pre-training}
         \label{fig:PT}
     \end{subfigure}
        \caption{NCF Training Loss Graph}
        \label{fig:NCF2}
\end{figure}

For the NCF model, pre-training GMF and MLP was also an essential hyperparameter in reducing the loss function and achieving better NDCG performance. As described in Section 3.3, the NCF model was constructed by concatenating the last hidden layer of GMF and MLP to learn the hidden features of user-item interactions. The concatenated layer is called the NeuMF layer. The training loss for the GMF and MLP are shown in Figure \ref{fig:GMF} and \ref{fig:MLP}. From Figure \ref{fig:GMF}, we can see that the binary cross-entropy loss for GMF starts at approximately 0.50 and drops down to 0.10 and from Figure \ref{fig:MLP}, it can be seen that the binary cross-entropy loss for MLP begins around at 0.35 and ends at 0.07. We trained the NCF model without pre-training weights and observed the training loss. From Figure \ref{fig:WPT}, we can see that the binary cross-entropy loss is very high, which starts at approximately 0.35. \cite{he2017neural} argued that the pre-trained weights reduced the training loss; we trained the GMF and MLP model separately with an Adam optimizer. The trained GMF and MLP model weights are used to train the NeuMF layer. As a result, the binary cross-entropy training loss reduced drastically and started at 0.064, which is shown in Figure \ref{fig:PT}, whereas the non-pre-trained model started at 0.35. The pre-trained model achieved 0.724 on the NDCG metric, while the non-pre-trained model achieved 0.698, a 2.6\% drop in NDCG performance, showing that pre-training has a significant impact on the training loss. For the ACF model, we implemented different loss functions like mean squared loss and binary cross entropy loss to see any improvement in the overall performance of the model. We found that changing the loss function did not improve the NDCG performance; we believe this is likely due to the sparse nature of the dataset. 


Hyperparameter tuning by Bayesian Optimization (BO) sometimes results in overfitting, the knowledge of the validation data split(s) increases as the number of iterations increases on optimizing hyperparameters. k-fold cross-validation is the standard solution to the problem of overfitting in hyperparameter optimization. The authors \cite{wainer2017empirical} conducted an experimental study to find the number of validation folds to be used for cross-validation evaluation, and they found that two or three validation folds are adequate for finding optimal hyperparameters. In this paper, we used NDCG as a target function and applied Bayesian Optimization to find the optimal combination of parameters that maximized the NDCG. We selected the iterations (number of validation folds) to be three, and the hyperparameters include the number of epochs, learning rate, batch size, weight decay, etc. Table \ref{tab:BO} shows the Bayesian Optimization results for the NCF model with two hyperparameters, namely learning rate and n\_factors. From Table \ref{tab:BO} it can be seen that learning rate 0.001 and n\_factors 16 gives the maximum NDCG (0.708). This discussion motivates further work on how to best represent sparse datasets to predict product recommendations efficiently. The CF with NCF predictions demonstrated in this paper discovers the hidden features of user-item interactions that lead to a good generalization, though a hybrid approach of combining item-based and user-based recommenders for predictions on a more complex representation of user-item interactions may improve the NDCG results presented in this paper. The extensions to matrix factorization and the usage of click data for future work are discussed in the next section.

\begin{table}[ht]
\caption{Bayesian Optimization for NCF}
\begin{center}
\begin{tabular}{ |c|c|c|c| } 
\hline
\textbf{Iterations} & \textbf{NDCG@12} & \textbf{Learning Rate} & \textbf{n\_factors} \\
\hline
1 & 0.698  & 0.003 & 12\\
\hline
2 & 0.703 & 0.09 & 14\\
\hline
3 & 0.708  & 0.001 & 16\\
\hline
\end{tabular}
\end{center}
\label{tab:BO}
\end{table}

\section{Future Work and Conclusions}

\subsection{Future Work}

Research into collaborative filtering with matrix factorization has demonstrated algorithmic, logical, and practical insights that explain the NDCG performance observed in this paper and has provided future directions of research in predicting sparse user-item interactions with implicit feedback by creating a similarity metric over the latent variable model \cite{borgs2017thy}. 

The cold start problem is an active area of research that also applies to the work in this paper. The cold start problem can be relevant to both users and items that have no reviews or history; in simple terms, what to recommend to users/items that are not part of a training dataset. Different procedures have been proposed to address the cold start problem. A user/item deep learning content-based recommender was proposed by \cite{volkovs2017dropoutnet} gives a probability of user/item interaction in the absence of particular user/item interaction in the training dataset. 

 Similar to ACF, which is implemented in this paper, VAEs also learn hidden non-linear representations of the data using a probabilistic framework \cite{kingma2013auto}. VAEs that implemented CF demonstrated state-of-the-art results for movie recommendations on Netflix dataset  \cite{liang2018variational}. 

Another critical area of future work is to utilize click-data. In this work, we did not have click-data. However, we suggested the Company to consider implementing click-data so that it can be incorporated in future versions of the Company’s recommender system. Clickthrough Rate
(CTR) is a ratio that indicates how often the people who
see the recommendation ends up clicking it. In
future, after accumulating click-data, the Deep Interest Network \cite{zhou2018deep} model could
be implemented with the current use case and investigated
further. 
\subsection{Conclusions}

In this paper, the problem of predicting sparse product recommendations at a specific customer level was studied on a proprietary dataset from August 2016 to June 2019. It was hypothesized that CF with latent variable models could discover hidden features in the broad range of products by exploiting the similarity of customers’ past purchasing behaviors, which could be generalized to future purchases. CF with NCF using latent hidden factors for customers and products presented the highest average performance metrics over the period of the dataset studied. The BPR model is shown to be nearly as efficient as CF with NCF. Deep Learning strategies for CF with latent variables like ACF were investigated and revealed to not be as powerful as NCF. A benefit of the NCF approach is that it has discovered that non-linear hidden relationships exist between customers and products, information that was very valuable in the product recommendations. A limitation of NCF is that the learned hidden representation of user/item matrices is not interpretable, which makes it complicated to comprehend why NCF recommends these products to a customer. Recent work illustrated that matrix factorization with latent variables not limited to low-rank approximation shows excellent generalization properties, enduring sparsity and predicting CTR using Deep Interest Network are promising for further application development in predicting product recommendations where data may be sparse and underlying factors may be complicated, noisy and challenging to apprehend explicitly.

% or


\ifCLASSOPTIONcompsoc
  \section*{Acknowledgments}
\else
  \section*{Acknowledgment}
\fi

The authors would like to thank the Center for Management of Technology and Entrepreneurship (CMTE) and their sponsors for providing funding to our project.

\ifCLASSOPTIONcaptionsoff
  \newpage
\fi



\bibliographystyle{IEEEtran}
\bibliography{paper.bib}
%\begin{thebibliography}{1}


%\end{thebibliography}


\begin{IEEEbiography}[{\includegraphics[width=1in,height=1.25in,clip,keepaspectratio]{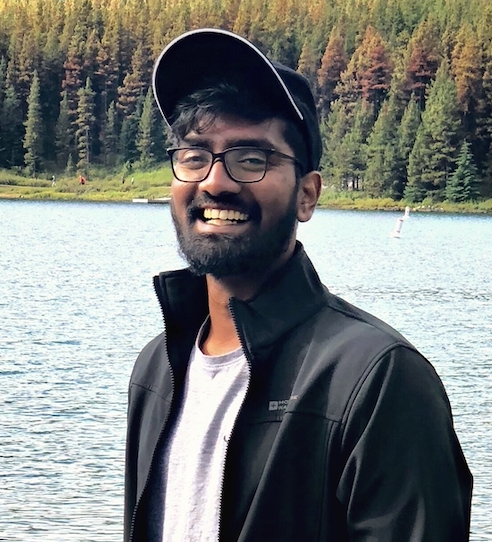}}]{Karthik Raja Kalaiselvi Bhaskar}

is currently working as a Student Researcher at Vector Institute advised by Prof. Bo Wang and an M.A.Sc graduate in the Department of Electrical and Computer Engineering at University of Toronto specialized in Machine Learning, advised by Prof. Deepa Kundur and by Prof. Yuri Lawryshyn. His current research focuses on at the intersection of  Machine Learning, Computational Biology, and Natural Language Processing.. Apart from my research, He is also interested in Deep Learning and Deep Reinforcement Learning with the focus of transfer learning, imitation learning, model-based RL. His ultimate goal is to build robust, privacy-preserved, and interpretable algorithms with human like ability to generalize in real-world environments by using data as its own supervision. Additional information can be found at: \url{https://www.comm.utoronto.ca/~kbhaskar/}.
\end{IEEEbiography}

\begin{IEEEbiography}[{\includegraphics[width=1in,height=1.05in,clip,keepaspectratio]{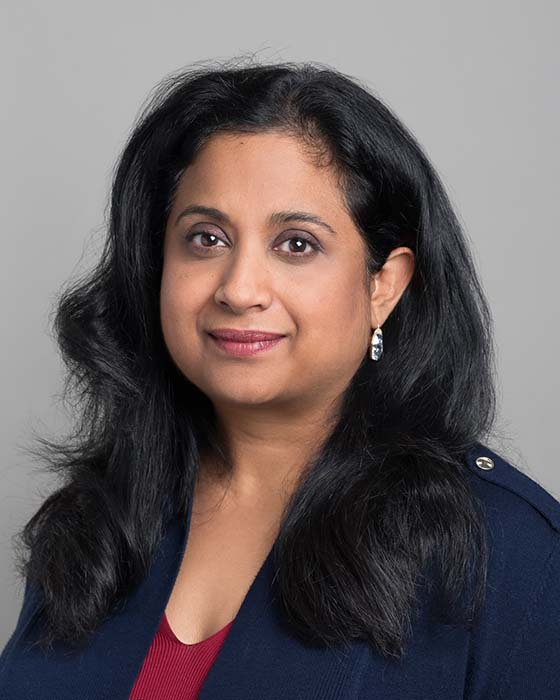}}]{Deepa Kundur}
is Professor \& Chair of The Edward S. Rogers Sr. Department of Electrical \& Computer Engineering at the University of Toronto. A native of Toronto, Canada, she received the B.A.Sc., M.A.Sc., and Ph.D. degrees all in Electrical and Computer Engineering in 1993, 1995, and 1999, respectively, from the University of Toronto. From January 2003 to December 2012 she was a faculty member in Electrical \& Computer Engineering at Texas A\&M University, and from September 1999 to December 2002 she was a faculty member in Electrical \& Computer Engineering at the University of Toronto. Professor Kundur’s research interests lie at the interface of cybersecurity, signal processing and complex dynamical networks. She is an author of over 200 journal and conference papers. She is also a recognized authority on cyber security issues and has appeared as an expert in popular television, radio and print media. Professor Kundur has participated on several editorial boards and currently serves on the Advisory Board of IEEE Spectrum. She  has served in numerous conference executive organization roles including as General Chair of the 2018 GlobalSIP Symposium on Information Processing, Learning and Optimization for Smart Energy Infrastructures, TPC Co-Chair for IEEE SmartGridComm 2018. Symposium Co-Chair for the Communications for the Smart Grid Track of ICC 2017, General Chair for the Workshop on Communications, Computation and Control for Resilient Smart Energy Systems at ACM e-Energy 2016, General Chair for the Workshop on Cyber-Physical Smart Grid Security and Resilience at Globecom 2016, General Chair for the Symposium on Signal  and Information Processing for Smart Grid Infrastructures at GlobalSIP 2016, General Chair for the 2015 International Conference on Smart Grids for Smart Cities , General Chair for the 2015 Smart Grid Resilience (SGR) Workshop at IEEE GLOBECOM 2015 and General Chair for the IEEE GlobalSIP’15 Symposium on Signal and Information Processing for Optimizing Future Energy Systems. Professor Kundur’s research has received best paper recognitions at numerous venues including the 2015 IEEE Smart Grid Communications Conference, the 2015 IEEE Electrical Power and Energy Conference, the 2012 IEEE Canadian Conference on Electrical \& Computer Engineering, the 2011 Cyber Security and Information Intelligence Research Workshop and the 2008 IEEE INFOCOM Workshop on Mission Critical Networks. She has also been the recipient of teaching awards at both the University of Toronto and Texas A\&M University. She is a Fellow of the IEEE, a Fellow of the Canadian Academy of Engineering, and a Senior Fellow of Massey College. Additional information can be found at: \url{https://www.ece.utoronto.ca/people/kundur-d/}.

\end{IEEEbiography}

\begin{IEEEbiography}[{\includegraphics[width=1in,height=1.25in,clip,keepaspectratio]{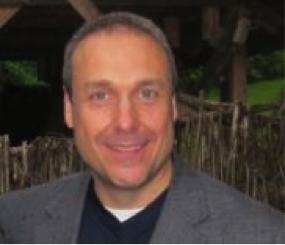}}]{Yuri Lawryshyn}
completed his Ph.D. degree in engineering at the University of Toronto specializing in Computational Fluid Dynamics (CFD), an MBA degree at the Richard Ivey Business School (UWO), and a Financial Engineering diploma at the Schulich School of Business (York U). Professor Lawryshyn have worked in a number of industries in research and development, management, marketing, and strategy. In 2007, Professor Lawryshyn joined the faculty of Chemical Engineering at University of Toronto. Professor Lawyrshyn’s research is focused on applying mathematical and numerical modeling in two broad areas: financial analysis and applied computational fluid dynamics (CFD). A key aspect of his research is the application of modeling to solve practical problems. His financial analysis research is primarily focused in the valuation of engineering projects with an emphasis on real options. Professor Lawyrshyn’s CFD research is primarily focused on water and wastewater treatment, and especially, UV disinfection. As Director of the Centre for Management of Technology and Entrepreneurship (CMTE), Professor Lawyrshyn is involved in a number of research projects with the CMTE’s corporate partners, namely, the Bank of Montreal (BMO), Canadian Imperial Bank of Commerce (CIBC) and Toronto Dominion (TD). Additional information can be found at: \url{http://www.labs.chem-eng.utoronto.ca/lawryshyn/}.
\end{IEEEbiography}

%\newpage


%\vfill


\end{document}